\documentclass[12pt,a4paper]{article}
\usepackage[dvips]{epsfig}
\title{
  {\vspace{-2cm} \normalsize
     \epsfig{figure=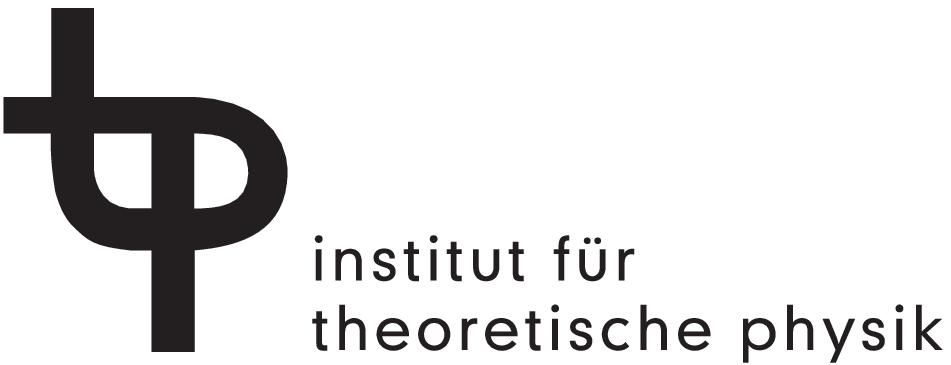,width=80mm}
     \hfill\parbox[b][30mm][t]{38mm}{MS-TP-02-13 \\
                                     cond-mat/0209201}  }\\[25mm]
The classical nucleation rate in two dimensions
}
\author{G.~M\"unster$^{\,(a)}$ and S.B.~Rutkevich$^{\,(b)}$
    \thanks{present address: Universit\"at Essen,
     Fachbereich 7 - Physik, Universit\"atsstr.\ 5, D-45117 Essen}\\
    {\small (a) Institut f\"ur Theoretische Physik,
        Universit\"at M\"unster,}\\
        {\small Wilhelm-Klemm-Str.~9, D-48149 M\"unster, Germany;}
        {\small e-mail: munsteg@uni-muenster.de}\\
    {\small (b) Institute of Physics of Solids and Semiconductors,}\\
        {\small P.Brovki 17, Minsk 220072, Belarus;}
        {\small e-mail: rut@ifttp.bas-net.by}}
\date{September 6, 2002}
\begin{document}
\maketitle

\begin{abstract}
In many systems in condensed matter physics and quantum field theory,
first order phase transitions are initiated by the nucleation of bubbles
of the stable phase.  In homogeneous nucleation theory the nucleation
rate $\Gamma$ can be written in the form of the Arrhenius law:
$\Gamma=\mathcal{A} e^{-\mathcal{H}_{c}}$.  Here $\mathcal{H}_{c}$ is
the energy of the critical bubble, and the prefactor $\mathcal{A}$ can
be expressed in terms of the determinant of the operator of fluctuations
near the critical bubble state.  In general it is not possible to find
explicit expressions for $\mathcal{A}$ and $\mathcal{H}_{c}$.  If the
difference $\eta$ between the energies of the stable and metastable
vacua is small, the constant $\mathcal{A}$ can be determined within the
leading approximation in $\eta$, which is an extension of the ``thin
wall approximation''.  We have done this calculation for the case of a
model with a real-valued order parameter in two dimensions.
\\[5mm]
\end{abstract}
%
\section{Introduction}

The problem of the decay of the metastable false vacuum at first order
phase transitions has attracted considerable interest due to its
numerous relations with condensed matter physics \cite{RG}, quantum
fields \cite{St}, cosmology \cite{Guth}, and black hole theory
\cite{Kastrup}.  In Langer's theory of homogeneous nucleation
\cite{Langer1,Langer2}, the false vacuum decay is associated with the
spontaneous nucleation of a critical bubble of a stable phase in a
metastable surrounding.  In the context of quantum field theory, the
nucleation theory was developed by Voloshin et al.\ \cite{Voloshin1},
and Callan and Coleman \cite{Col,CC}.  The quantity of main interest is
the nucleation rate $I$ per time and volume.

The nucleation rate
\begin{equation}
I = \frac{\kappa}{2\pi}\, \Gamma
\end{equation}
is a product of the static part $\Gamma$ and the so-called kinetic
prefactor $\kappa$, which depends on the detailed non-equilibrium
dynamics of the model, see \cite{Langer2,RG}. Most important is the
static nucleation rate $\Gamma$, which is equal to twice the imaginary
part of the free energy density of the metastable phase. In this
article we study the static part $\Gamma$.

In the homogeneous nucleation theory it has the form of the Arrhenius
law:
\begin{equation}
\Gamma = \mathcal{A}\exp(-\mathcal{H}_{c}),
\label{Arr}
\end{equation}
where $\mathcal{H}_{c}$ is the energy of the critical bubble.  The
prefactor $\mathcal{A}$ is determined by fluctuations near the critical
bubble state and can be expressed in terms of the functional determinant
of the fluctuation operator \cite{Langer1,CC}.

In the general case, it is not possible to find the explicit critical
bubble solution of the field equations analytically.  However, the
problem becomes asymptotically solvable, if the decaying metastable
state is close enough in energy to the stable one, i.e.\ if the energy
density difference $\eta$ between the metastable and stable vacua is
small.  The leading approximation in this small parameter is usually
called the ``thin wall approximation'' \cite{Linde}, since at $\eta
\rightarrow 0$ the critical bubble radius goes to infinity and becomes
much larger than the thickness of the bubble wall.

In the thin wall approximation, the critical bubble energy
$\mathcal{H}_{c}$ can be easily obtained from Langer's nucleation
theory.  It turns out to be much more difficult to find explicitly the
prefactor $\mathcal{A}$ in Eq.~(\ref{Arr}).  This problem, which is
important for applications of nucleation theory, has been extensively
studied in different models.

A remarkable result on this subject was obtained by Voloshin
\cite{Voloshin2}.  He considered scalar field theory in 2 dimensions
with a potential $U(\phi)$ of the type shown in Fig.~1.

\begin{figure}[hbt]
\vspace{.8cm}
\centering
\epsfig{file=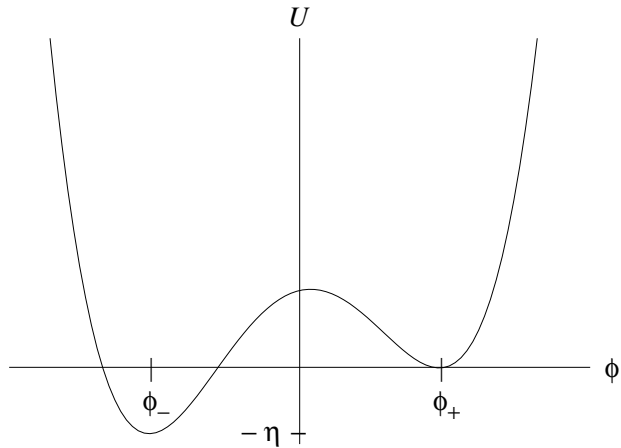,width=9cm}
\parbox[t]{0.8\textwidth}{
\caption{The potential $U$ with the false $(\phi_{+})$ and true
$(\phi_{-})$ vacuum}
}
\end{figure}

Voloshin claimed that in the limit $\eta \rightarrow 0$ the nucleation
rate $\Gamma$ in such a model can be described by the simple universal
formula
\begin{equation}
\Gamma =
\frac{\eta}{2\pi} \exp\left(-\frac{\pi\sigma^{2}}{\eta} \right).
\label{V}
\end{equation}
Here $\sigma$ is the surface tension of the wall between the stable and
metastable vacua in the limit $\eta \rightarrow 0.$ Thus, according to
\cite {Voloshin2}, in this limit the nucleation rate $\Gamma$ is
determined by two well defined macroscopic parameters $\eta$ and
$\sigma$.  Another claim of \cite{Voloshin2} is that there are no
corrections to Eq.~(\ref{V}) proportional to powers
of the dimensionless parameter $\eta/\sigma^{2}$.  Voloshin arrived at
these conclusions by an analysis performed in the thin wall
approximation.  He replaced the original scalar field theory by an
effective geometrical one, which describes only fluctuations of the
critical bubble shape.  This approach implies that all other
fluctuations of the original scalar field could be properly accounted
for by the correct choice of the macroscopic parameters $\eta$ and
$\sigma$.

Recently an analytical method was developed \cite{MR}, which allows one
to study nucleation in the scalar field model beyond the thin wall
approximation.  In \cite{MR} this method was used to calculate the
nucleation rate for the first order phase transition in the
three-dimensional Ginzburg-Landau model.  In the present paper we apply
the same approach to the two-dimensional case.  We calculate the
nucleation rate beyond the thin wall approximation and verify directly
Voloshin's claim (\ref{V}).

Nucleation theory in two-dimensional scalar field theory has also been
studied by Kiselev and Selivanov \cite{Kiselev}, Strumia and Tetradis
\cite {ST}, and other authors.  In these articles, however, different
renormalization schemes have been used and $\Gamma$ has not been
expressed in terms of macroscopic parameters $\eta$ and $\sigma$.  This
makes it difficult to compare their results with the ones discussed in
this article.

In the articles \cite{R1,R2} the nucleation rate was calculated in the
two-dimensional Ising model in a small magnetic field for arbitrary
anisotropies. If Voloshin's result (\ref{V}) is universal, it should be
applicable as well to the Ising model in the critical region.  Indeed,
the results of \cite{R1,R2}, rewritten in terms of $\eta$ and $\sigma$,
are in a very good agreement with Eq.~(\ref{V}).  The exponent factors
are the same, and the prefactors differ only by the number
$\pi^{2}/9\approx 1.0966$, which is very close to unity.  This small
discrepancy increased our interest in the subject of the present study.

%
%
\section{Model and notations}

We consider the two-dimensional asymmetric Ginzburg-Landau model defined
by the Hamiltonian:
\begin{equation}
\mathcal{H}(\phi) = \int\!\!d^{2}x\,\left[\frac{1}{2}
(\partial_{\mu}\,\phi(x))^{2} + U(\phi(x))\right],
\end{equation}
where $\phi(x)$ is the continuous one-component order parameter, and the
potential $U(\phi)$ depicted in Fig.~1 is given by
\begin{equation}
U(\phi) = U_{s}(\phi) + \frac{\eta_{0}}{2\,v} (\phi-v) + U_{0}.
\label{U}
\end{equation}
Here $U_{s}(\phi)$ denotes the symmetric part of the potential:
\begin{equation}
U_{s}(\phi) = \frac{g}{4!} \left(\phi^{2} - v^{2}\right)^{2}.
\end{equation}
The potential $U(\phi)$ has a metastable minimum (false vacuum) at
$\phi=\phi_{+}$ and a stable one (true vacuum) at $\phi=\phi_{-}$. The
constant term $U_{0}$ in Eq.~(\ref{U}) is chosen to ensure
$U(\phi_{+})=0$.

The partition function is given by the functional integral
\begin{equation}
Z = \int\!\!D\phi\,\exp\left[-\mathcal{H}(\phi)\right].
\end{equation}
The temperature has been absorbed into $\mathcal{H}$.

It is convenient to define the mass $m$ and the inverse coupling
parameter $\beta$ by
\begin{equation}
m^{2} = \frac{\partial^{2}}{\partial\phi^{2}} U_{s}(\phi)\mid_{\phi=v}
= \frac{g\,v^{2}}{3},\qquad
\beta = \frac{3m^{2}}{g},
\end{equation}
and to introduce dimensionless quantities
\begin{equation}
\tilde{x}_{\mu} = \frac{m}{2}\,x_{\mu},\quad
\tilde{\eta} = \frac{g}{2\,m^{4}}\,\eta_{0},\quad
\varphi(\tilde{x}) = \frac{\phi(x)}{v},\quad
\varphi_{\pm} = \frac{\phi_{\pm}}{v},\quad
\tilde{\mathcal{H}} = \frac{\mathcal{H}}{\beta}\,.
\end{equation}
In dimensionless variables the Hamiltonian and partition function take
the form
\begin{equation}
\tilde{\mathcal{H}}(\varphi) = \int\!\!d^{2}\tilde{x}\;
\left[\frac{1}{2}\left(\nabla\varphi\right)^{2}
+ \tilde{U}\left(\varphi(\tilde{x})\right)\right],
\end{equation}
where
\begin{equation}
\tilde{U}(\varphi) = \frac{1}{2} \left[\left(\varphi^{2} - 1\right)^{2}
- \left(\varphi_{+}^{2} - 1\right)^{2}\right]
+ \frac{4}{3}\,\tilde{\eta} \left(\varphi - \varphi_{+}\right),
\end{equation}
and
\begin{equation}
Z = \int\!\!D\varphi(\tilde{x})\,
\exp\left[-\beta \tilde{\mathcal{H}}(\varphi)\right].
\end{equation}
%
%
\section{The critical bubble solution}

The uniform solutions of the field equation
\begin{equation}
\delta\tilde{\mathcal{H}}/\delta\varphi(\tilde{x})=0
\label{vac}
\end{equation}
are the stable $\varphi_{-}$ and false (metastable) $\varphi_{+}$ vacua
given by
\begin{equation}
\varphi_{\pm} =
\pm 1 - \frac{\tilde{\eta}}{3} \mp \frac{\tilde{\eta}^{2}}{6}
- \frac{4\tilde{\eta}^{3}}{27} + O(\tilde{\eta}^{4}).
\label{fipm}
\end{equation}
The critical bubble $\varphi_{b}(\tilde{x})$ is the non-uniform radially
symmetric solution of Eq.~(\ref{vac}) approaching the false vacuum at
infinity. That is,
\begin{equation}
-\frac{d^{2}\varphi_{b}}{d\,\tilde{r}^{2}}
-\frac{1}{\tilde{r}}\frac{d\varphi_{b}}{d\,\tilde{r}}
+ 2\varphi_{b}(\varphi_{b}^{2} - 1)
+ \frac{4}{3}\,\tilde{\eta} = 0, \qquad \qquad
\lim_{\tilde{r}\rightarrow \infty} \varphi_{b}(\tilde{r}) = \varphi_{+},
\label{bub}
\end{equation}
where $\tilde{r} = \sqrt{\tilde{x}_{\mu}\,\tilde{x}_{\mu}}.$ The profile
of the critical bubble solution is shown schematically in Fig.~2. If
$\tilde{\eta}$ is small, the thin wall centered at $\tilde{r}=\tilde{R}$
divides regions of false and stable vacua outside and inside the bubble,
respectively.

\begin{figure}[hbt]
\vspace{.8cm}
\centering
\epsfig{file=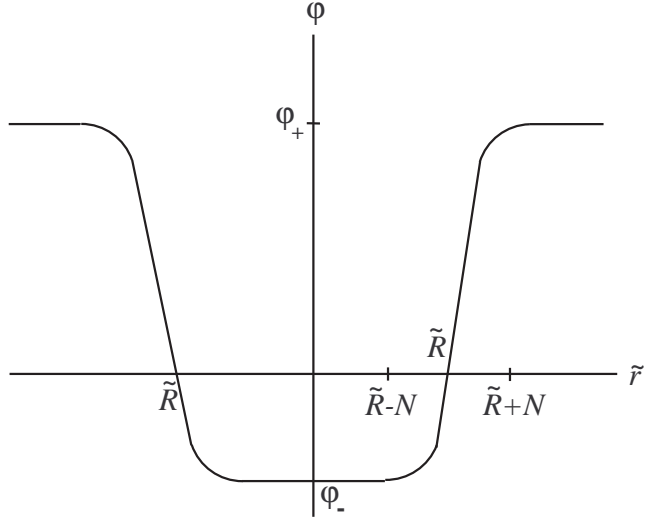,width=9cm}
\parbox[t]{0.8\textwidth}{
\caption{Profile of the critical bubble}
}
\end{figure}

Equation (\ref{bub}) can not be solved explicitly.  Following the
approach introduced in \cite{MR} we shall construct the solution by
expansion in powers of $\tilde{\eta}$. Introducing the new independent
variable $\xi$:
\begin{equation}
\xi = \tilde{r} - \tilde{R},
\label{ksi}
\end{equation}
we expand $\tilde{R}$ and $\varphi_{b}(\xi )$ as
\begin{eqnarray}
\tilde{R} &=& \frac{a_{-1}}{\tilde{\eta}}+a_{0} + a_{1}\tilde{\eta}
+ a_{2}\tilde{\eta}^{2} + O(\tilde{\eta}^{3}),\label{R1}\\
\varphi_{b}(\xi) &=& \varphi_{0}(\xi) + \varphi_{1}(\xi)\,\tilde{\eta}
+ \varphi_{2}(\xi)\tilde{\eta}^{2} + O(\tilde{\eta}^{3}).
\label{fi1}
\end{eqnarray}
After substitution of Eqs.~(\ref{ksi}--\ref{fi1}) into (\ref{bub}) one
obtains perturbatively in $\tilde{\eta}$:
\begin{eqnarray}
a_{-1} &=&\frac{1}{2},\;a_{0}=0,\;a_{1}=-\frac{2}{9},\;a_{2}=0,
\label{asy}\\
\varphi_{0}(\xi) &=& \tanh\xi,\qquad
\varphi_{1}(\xi) = -\frac{1}{3},\nonumber\\
\varphi_{2}(\xi) &=& -\frac{1}{24\cosh^{2}\xi}\,
\biggl\{10\xi - 16\xi\cosh(2\xi) - 2\xi\cosh(4\xi) + \nonumber\\
&&2\ln\left[2\cosh\xi\right]\left[12\xi + 8\sinh(2\xi) + \sinh(4\xi)
\right] - 24\int\limits_{0}^{\xi}\!\!dt\,\,t\tanh t\biggl\}. \nonumber
\end{eqnarray}
The bubble energy
$\tilde{E} = \tilde{\mathcal{H}}\left[\varphi_{b}(x)\right]$
can be written as
\begin{equation}
\tilde{E} =
\pi\int\limits_{-\tilde{R}}^{\infty}\!\!d\xi\,(\tilde{R}+\xi)
\left(\frac{d\varphi_{b}(\xi)}{d\xi}\right)^{2}.
\label{ener}
\end{equation}
Substitution of Eq.~(\ref{asy}) into (\ref{ener}) yields
\begin{equation}
\tilde{E} = \frac{2\pi}{3} \left[\frac{1}{\tilde{\eta}}
+ \tilde{\eta} \left(\frac{19}{18} - \frac{\pi^{2}}{3}\right)
+ O(\tilde{\eta}^{3})\right].
\label{E}
\end{equation}

It is the basic principle of homogeneous nucleation theory that the
decay of the metastable vacuum occurs through nucleation of the critical
bubble. Callan and Coleman expressed the nucleation rate $\Gamma$ of
the metastable vacuum in terms of functional determinants \cite{Col,CC}.
In our notation their result takes the form
\begin{equation}
\tilde{\Gamma} = \frac{\beta\,\tilde{E}}{2\pi}
\frac{1}{\sqrt{\left|\lambda_{0}\right|}}
\exp\left(-\beta\,\tilde{E} + S\right).
\label{W}
\end{equation}
Here $\tilde{\Gamma} = 4\Gamma/m^{2}$ is the dimensionless nucleation
rate, and the entropy $S$ associated with the critical bubble is given
by
\begin{equation}
\exp S = \left[\frac{\det^{\prime} M}{\det M^{(0)}}\right]^{-1/2},
\label{S}
\end{equation}
where $M$ and $M^{(0)}$ are the fluctuation operators near the bubble
$\varphi_{b}(\tilde{x})$ and the metastable uniform vacuum
$\varphi_{+}$, respectively:
\begin{eqnarray}
M &=& \mathcal{-\partial}^{2}
+6\,\left[\varphi_{b}(\tilde{r})\right]^{2} - 2, \\
M^{(0)} &=& \mathcal{-\partial}^{2} + 6\,\varphi_{+}^{2} - 2.
\end{eqnarray}
The operator $M$ has two zero modes proportional to
$\mathcal{\partial}_{\mu}\,\varphi_{b}(\tilde{x})$, $\mu =1,2,$ and one
negative mode with the eigenvalue
\begin{equation}
\lambda_{0}=-4\,\tilde{\eta}^{2}.
\label{lam0}
\end{equation}
The notation $\det^{\prime}$ implies that the three above mentioned
modes are omitted in the corresponding determinant. After substitution
of Eqs.~(\ref{E}) and (\ref{lam0}), equation (\ref{W}) simplifies to
\begin{equation}
\tilde{\Gamma} = \frac{\beta}{6\,\tilde{\eta}^{2}}
\exp\left(-\,\frac{2\pi\beta}{3\,\tilde{\eta}} + S\right)
\left(1 + O(\tilde{\eta})\right).
\end{equation}
In the subsequent sections we shall calculate the small $\tilde{\eta}$
expansion for the critical bubble entropy (\ref{S}) to the order
$O(\tilde{\eta}^{0})$.
%
%
\section{The bubble entropy}

The spectrum of the fluctuation operator $M$ can be determined in the
form of a perturbative expansion in powers of the parameter
$\tilde{\eta}$ as in \cite{MR}.  This is achieved in the following way.

Introducing the angular momentum quantum number $\mu \in \mathbf{Z}$ in
two dimensions, the radial Schr\"o\-din\-ger operators corresponding to
$M$ and $M^{(0)}$ are
\begin{eqnarray}
H_{\mu} &=&-\frac{d^{2}}{d\,\tilde{r}^{2}}
- \frac{1}{\tilde{r}} \frac{d}{d\,\tilde{r}}
+ \frac{\mu^{2}}{\tilde{r}^{2}}
+ 6\ \left[ \varphi_{b}(\tilde{r})\right]^{2} - 2, \\
H_{\mu }^{(0)} &=&-\frac{d^{2}}{d\,\tilde{r}^{2}}
- \frac{1}{\tilde{r}} \frac{d}{d\,\tilde{r}}
+ \frac{\mu^{2}}{\tilde{r}^{2}}
+ 6\ \varphi_{+}^{2} - 2.
\end{eqnarray}
Shifting the coordinate from $\tilde{r}$ to $\xi$ and making use of the
Laurent series (\ref{R1}) for $\tilde{R}$, the eigenvalue problem for
$H_{\mu}$ can be treated perturbatively in $\tilde{\eta}$.
The lowest order leads to the exactly solvable P\"oschl-Teller operator
\begin{equation}
-\frac{d^{2}}{d \xi^{2}} - 6\, \textrm{sech}^2 \xi + 4,
\end{equation}
which has discrete eigenvalues 0 and 3 and a continuum above 4. In
second order one finds a band around 0:
\begin{equation}
\lambda_{0 \mu} = 4 \tilde{\eta}^2 ( \mu^2 - 1) + O( \tilde{\eta}^4 ),
\end{equation}
a band around 3:
\begin{equation}
\lambda_{3 \mu} = 3 + 4 \tilde{\eta}^2 ( \mu^2 + \textrm{const.})
+ O( \tilde{\eta}^4 ),
\end{equation}
and a continuum
\begin{equation}
\lambda_{k \mu} = k^2 + 6 \varphi_{+}^2 - 2 + 4 \tilde{\eta}^2 \mu^2
+ O( \tilde{\eta}^4 )
\end{equation}
with $k \in \mathbf{R}$.
This spectrum includes the negative mode $\lambda_{00}$ and the two zero
modes $\lambda_{0,\pm 1}$.

The sum over $\mu$ and the integration over $k$ produce ultraviolet
divergencies in $S$.  We treat these by means of dimensional
regularization in $d=2-\varepsilon$ dimensions.  As there appear volume
integrals in intermediate steps of the calculation, the extra dimensions
are equipped with a finite extent $L$ and periodic boundary conditions.
The parameter $L$ must cancel out in finite results.

The finite part of the regularized entropy can be conveniently evaluated
with the help of zeta-function techniques \cite{M4}.  The operator-zeta
function appropriate for our case is defined by
\begin{equation}
\zeta_{M}(z) = \frac{1}{\Gamma(z)}
\int_{0}^{\infty}\!\!dt\,t^{z-1}\,\left(\textrm{Tr}\,' e^{-tM} -
\textrm{Tr}\, e^{-t M^{(0)}} \right)
\end{equation}
for $\textrm{Re}\,z > 1$ and analytical continuation to other values of
$z$. The integrand contains the heat kernels $\exp (-tM)$ and $\exp (-t
M^{(0)})$.  For positive $t$ there is an asymptotic expansion, the
so-called Seeley expansion, which is of the form
\begin{equation}
\textrm{Tr} \left( e^{-tM} - e^{-t M^{(0)}} \right)
= (4\pi t)^{- d/2} \sum_{n=1}^{\infty} t^n\, \mathcal{O}_n\,.
\end{equation}
Following \cite{M4} one obtains
\begin{equation}
S = \frac{1}{2}\,\frac{d}{dz} \zeta_{M} (0)
+ \frac{\mathcal{O}_1}{8\pi} \left[ \frac{2}{\varepsilon} + \ln 4\pi
+ \Gamma' (1) \right] + O(\varepsilon).
\end{equation}
This expression displays the divergence as a simple pole in
$\varepsilon$. The derivative of the zeta-function is a finite quantity.
The first Seeley coefficient is given by
\begin{equation}
\mathcal{O}_1 =
-6 \int\!\!d^d \tilde{x}\, ( [ \varphi_b (\tilde{r})]^2
- \varphi_{+}^2 )
= \tilde{L}^{- \varepsilon} \frac{10\pi}{\tilde{\eta}}
+ O(\tilde{\eta}),
\end{equation}
where
\begin{equation}
\tilde{L} = \frac{m}{2} L.
\end{equation}

The zeta-function is decomposed into a contribution from the band near
zero and the rest,
\begin{equation}
\zeta_{M} (z) = \zeta_0 (z) + \zeta_1 (z),
\end{equation}
where
\begin{equation}
\zeta_0 (z) = \frac{1}{\Gamma(z)} \int_{0}^{\infty}\!\!dt\,t^{z-1}\,
\sum_{\mu \neq 0,\pm 1} e^{- t \lambda_{0 \mu}}.
\end{equation}
Correspondingly, the entropy is decomposed as
\begin{equation}
S = S_0 + S_1
+ \tilde{L}^{-\varepsilon} \frac{5}{4 \tilde{\eta}}
\left[ \frac{2}{\varepsilon} + \ln 4\pi + \Gamma' (1) \right]
+ O(\tilde{\eta}) + O(\varepsilon).
\label{Sdiv}
\end{equation}
Consider the part
\begin{equation}
S_0 = \frac{1}{2} \frac{d}{dz} \zeta_0 (0).
\end{equation}
For a general spectrum of the type
\begin{equation}
\lambda_{\mu} = a (\mu + b)(\mu + c)
\end{equation}
the logarithm of the zeta-function regularized determinant is given by
\begin{equation}
- \frac{d}{dz} \zeta_0 (0) = - 2 \ln \Gamma (b+1) - 2 \ln \Gamma (c+1)
- (b+c) \ln a + \ln (bc) + 2 \ln (2\pi),
\end{equation}
which can be derived with the help of Riemann's and Hurwitz's
zeta-functions. Setting $a = 4 \tilde{\eta}^2, b=0, c=2$ one finds
\begin{equation}
S_0 = \ln \left( \frac{8 \tilde{\eta}^3 }{\pi} \right)
+ O(\tilde{\eta}^{2}).
\end{equation}
This piece will therefore contribute the factor
\begin{equation}
e^{S_0} =  \frac{8 \tilde{\eta}^3 }{\pi}
\{ 1 + O(\tilde{\eta}^{2})\}
\label{S0}
\end{equation}
to the prefactor $\mathcal{A}$ of $\Gamma$.

The remaining part $S_1$ of the entropy is calculated with the help of
methods from quantum mechanical scattering theory. The heat kernels can
be represented as
\begin{equation}
K_{t}(M) \equiv \textrm{Tr} \left( e^{-tM} - e^{-t M^{(0)}} \right)
= -\int\limits_{C}\frac{d\lambda}{2\pi i}\ e^{-\lambda t}\
\textrm{Tr}\left[ (\lambda - M)^{-1} - (\lambda - M^{(0)})^{-1}
\right],
\end{equation}
where the integration path $C$ in the complex plane is shown in Fig.~3
\begin{figure}[hbt]
\vspace{.8cm}
\centering
\epsfig{file=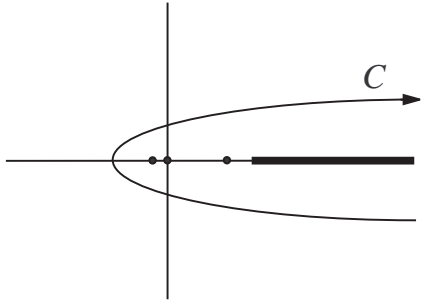,width=6cm}
\parbox[t]{0.8\textwidth}{
\caption{Integration path $C$ in the complex $\lambda$-plane}
}
\end{figure}

Decomposed into the angular momentum sums this reads
\begin{equation}
K_{t}(M)
= - \sum\limits_{\mu} \int\limits_{C}\frac{d\lambda}{2\pi i}\
e^{-\lambda t} A(\lambda,\mu),
\end{equation}
with
\begin{equation}
A(\lambda,\mu) =
\textrm{Tr}\left[ (\lambda - H_{\mu})^{-1} -
(\lambda - H_{\mu}^{(0)})^{-1} \right].
\end{equation}
We obtained an exact representation for $A(\lambda,\mu)$.
To describe it some notations are necessary.

Let $f_{i}(\tilde{r},\lambda),\ g_{i}(\tilde{r},\lambda),\ i=1,2$ be the
solutions of the linear ordinary differential equation
\begin{equation}
H_{\mu}\ \psi(\tilde{r}) = \lambda\ \psi(\tilde{r})
\label{dif}
\end{equation}
determined by their asymptotics:
\begin{eqnarray}
f_{1}(\tilde{r},\lambda) &\rightarrow &K_{\mu}(q\tilde{r}), \ \mbox{and}\
f_{2}(\tilde{r},\lambda)\rightarrow I_{\mu}(q\tilde{r}), \ \mbox{at}\
\tilde{r}\rightarrow \infty, \\
g_{1}(\tilde{r},\lambda) &\rightarrow &K_{\mu}(q_{-}\,\tilde{r}),
\ \mbox{and}\
g_{2}(\tilde{r},\lambda)\rightarrow I_{\mu}(q_{-}\,\tilde{r}),
\ \mbox{at}\ \tilde{r}\rightarrow 0.
\end{eqnarray}
Here $K_{\mu}(z)$ and $I_{\mu}(z)$ are modified Bessel functions, and
the parameters $q$ and $q_{-}$ are defined as
\begin{equation}
q = \left(6\varphi_{+}^{2} - \lambda - 2\right)^{1/2},\quad
q_{-} = \left(6\varphi_{-}^{2} - \lambda - 2\right)^{1/2}.
\end{equation}
Since the second order equation (\ref{dif}) has two linearly independent
solutions, there is a linear dependence between the functions
$g_{i}(\tilde{r},\lambda)$ and $f_{i}(\tilde{r},\lambda)$:
\begin{equation}
g_{i}(\tilde{r},\lambda) = \sum\limits_{j=1,2} \alpha_{ij}(\lambda)\
f_{j}(\tilde{r},\lambda).
\label{gi}
\end{equation}
The function $A(\lambda,\mu)$ can be expressed explicitly in terms of
the coefficient $\alpha_{22}(\lambda)$:
\begin{equation}
A(\lambda,\mu) = \frac{3\mu(\varphi_{+}^{2} - \varphi_{-}^{2})}
{(\lambda + 2 -6\varphi_{-}^{2})(\lambda + 2 - 6\varphi_{+}^{2})}
+ \frac{d\ln\alpha_{22}(\lambda)}{d\lambda}.
\label{tr}
\end{equation}

This representation for the trace of resolvent operators is
exact. However, equation (\ref{dif}) can not be solved in closed form
for arbitrary $\tilde{\eta}$. So we have to consider the
small-$\tilde{\eta}$ expansion for $\alpha_{22}(\lambda)$. We have
obtained two terms of this expansion by use of a perturbation
theoretical analysis of the scattering problem (\ref{dif}--\ref{gi}).
Omitting the details, the logarithmic derivative of the matrix element
$\alpha_{22}(\lambda)$ up to quadratic terms in $\tilde{\eta}$ takes
the form:
\begin{equation}
\frac{d\ln\alpha_{22}(\lambda)}{d\lambda} =
\frac{1}{(4 + p^{2} - \lambda)^{1/2}}
\left[\frac{2\,p^{2}}{\left(\lambda - 4\right)^{2}}
- \frac{1}{\lambda-4} + \frac{1}{\lambda - 3 - p^{2}}
+ \frac{2}{\lambda - p^{2}}\right]
+ O(\tilde{\eta}^{2}).
\label{a22}
\end{equation}
Here $p$ is the angular momentum parameter defined as
$p=2\,\tilde{\eta}\mu \approx \mu /\tilde{R}$.

Substitution of Eqs.~(\ref{a22}) and (\ref{fipm}) into (\ref{tr})
yields:
\begin{eqnarray}
A(\lambda,\mu) &=& \frac{1}{\left(4 + p^{2} - \lambda\right)^{1/2}}
\left[\frac{2\,p^{2}}{\left(\lambda - 4\right)^{2}}
- \frac{1}{\lambda-4} + \frac{1}{\lambda-3-p^{2}} +
\frac{2}{\lambda-p^{2}}\right]\nonumber\\
&&- \frac{2\,\left|p\right|}{\left(\lambda - 4\right)^{2}}
+ O(\tilde{\eta}^{2}).
\end{eqnarray}
This function has simple poles at $\lambda = p^2, \lambda = 3 + p^2$ and
a square root branching at $\lambda = 4 + p^2$. It is analytic at
$\lambda = 4$.

With the help of this expression we could evaluate the integral
representation for $K_{t}(M)$ and the related zeta-function. The details
of this lengthy analysis will not be presented here. The
$\mu$-summations have been done by means of Poisson's summation formula.
Separating the contribution of the band near zero, which has been
treated above, the final result for the remaining entropy is
\begin{equation}
S_1 = \frac{1}{2 \tilde{\eta}} (6 + \frac{2\pi}{\sqrt{3}} - 5\ln 2 ).
\label{S1}
\end{equation}
This is the central result of this section.
%
%
\section{Decay rate}

In $d=2-\varepsilon$ dimensions the bubble energy is associated with an
additional factor $L^{-\varepsilon}$ and the regularized dimensionless
decay rate $\tilde{\Gamma}$ is given by
\begin{equation}
\tilde{\Gamma}=
\frac{\beta L^{-\varepsilon}}{6\,\tilde{\eta}^{2}}
\exp\left(-\,\frac{2\pi\beta L^{-\varepsilon}}{3\,\tilde{\eta}}+S\right)
\left(1+O(\tilde{\eta})\right).
\end{equation}
In terms of the dimensionless quartic coupling
\begin{equation}
u = \frac{g}{m^{4-d}}
\end{equation}
the parameter $\beta$ is equal to
\begin{equation}
\beta = \frac{3}{u} m^{-\varepsilon}\,.
\end{equation}
Using the results (\ref{Sdiv},\ref{S0}), we can write the nucleation
rate as
\begin{equation}
\tilde{\Gamma}=
(2 \tilde{L}^{-\varepsilon}) \frac{4 \tilde{\eta}}{\pi u}
\exp\left(
- \frac{2\pi}{u \tilde{\eta}}\, (2 \tilde{L})^{-\varepsilon}
+ \tilde{L}^{-\varepsilon} \frac{5}{4 \tilde{\eta}}
\left[ \frac{2}{\varepsilon} + \ln 4\pi + \Gamma' (1) \right]
+ S_1 + O(\tilde{\eta}) + O(\varepsilon)
\right).
\end{equation}
The entropy contains an UV-divergent term, represented by a pole in
$\varepsilon$. After renormalization of the parameters of the model
according to the usual prescriptions, the divergencies as well as the
spurious $L$-dependence should disappear in the limit $\varepsilon \to
0$.

For the renormalization of the model parameters we use the same scheme
as in \cite{MR,M4}. A straightforward calculation yields the relation
between the bare and renormalized dimensionless couplings and masses on
the one-loop level:
\begin{equation}
u = u_R \left\{ 1 - \frac{u_R}{4\pi} \left[
\frac{2}{\varepsilon} + \ln 4\pi + \Gamma' (1) + \frac{3}{4} \right]
+ O(\varepsilon) + O(u_R^2)\right\}.
\end{equation}
\begin{equation}
m^2 = m_R^2 \left\{ 1 + \frac{u_R}{4\pi} \left[
\frac{2}{\varepsilon} + \ln 4\pi + \Gamma' (1) + \frac{7}{4} \right]
+ O(\varepsilon) + O(u_R^2)\right\}.
\end{equation}
The asymmetry parameter $\eta_{0}$ is renormalized as follows.
On tree level the difference between the minima of the potential
\begin{equation}
\Delta U = U(\phi_{+}) - U(\phi_{-}) = \eta_{0} + O(\eta_{0}^3)
\end{equation}
is equal to $\eta_{0}$ for small asymmetries. Let
$U_{\textrm{\scriptsize eff}}$ be the full effective potential given by
\begin{equation}
\Gamma[ \phi ] = - \int\!\!d^2 x\ U_{\textrm{\scriptsize eff}}(\phi)
\qquad \textrm{for} \ \phi = \textrm{const.},
\end{equation}
where $\Gamma[ \phi ]$ is the generating functional of one-particle
irreducible vertex functions. Then we define the renormalized asymmetry
parameter $\eta$ through
\begin{equation}
\eta = \Delta U_{\textrm{\scriptsize eff}} =
U_{\textrm{\scriptsize eff}} (\langle \phi \rangle_{+}) -
U_{\textrm{\scriptsize eff}} (\langle \phi \rangle_{-}).
\end{equation}
It is related to $\eta_{0}$ by
\begin{equation}
\eta = \frac{\eta_{0}}{v} \langle \phi \rangle + O(\eta_{0}^2),
\end{equation}
where $\langle \phi \rangle$ is the expectation value of the field at
$\eta_{0}=0$. From a one-loop calculation we get
\begin{equation}
\eta_{0} = \eta
\left\{ 1 + \frac{u_R}{8\pi} \left[
\frac{2}{\varepsilon} + \ln 4\pi + \Gamma' (1) \right]
+ O(\varepsilon) + O(u_R^2)\right\}.
\end{equation}
Expressing the unrenormalized parameters in terms of their renormalized
counterparts, the divergencies cancel indeed and in the limit
$\varepsilon = 0$ we obtain
\begin{equation}
\Gamma = \frac{\eta}{2\pi}\, e^{-F}
\end{equation}
with
\begin{eqnarray}
F &=& - S_1 + 4 \pi \frac{m_R^2}{u_R^2 \eta}
\left\{ 1 + \frac{u_R}{4\pi} \left(\frac{5}{4} - 5 \ln 2 \right)
+ O(u_R^2) \right\}
+ O(\eta)\\
&=& 4 \pi \frac{m_R^2}{u_R^2 \eta}
\left\{ 1 - \frac{u_R}{4\pi} \left( \frac{19}{4} - \frac{\pi}{\sqrt{3}}
\right) + O(u_R^2) \right\} + O(\eta).
\end{eqnarray}

To make contact with Voloshin's proposal we have to express this result
in terms of the interface tension $\sigma$. To this end we calculated
$\sigma$ along the lines of \cite{M4,M3}, but now for $d=2$.
Leaving out the details here, we obtained
\begin{equation}
\sigma = \frac{2 m_R}{u_R}
\left\{ 1 - \frac{u_R}{8\pi} \left( \frac{19}{4} - \frac{\pi}{\sqrt{3}}
\right) + O(u_R^2) \right\}.
\end{equation}
This implies
\begin{equation}
F = \frac{\pi \sigma^2}{\eta} \left\{ 1 + O(u_R^2) \right\}
+ O(\eta).
\end{equation}
Our final result is therefore
\begin{equation}
\Gamma = \frac{\eta}{2\pi}
\exp \left( - \frac{\pi \sigma^2}{\eta} \left\{ 1 + O(u_R^2) \right\}
+ O(\eta) \right).
\label{Gamma}
\end{equation}
This is in perfect agreement with Voloshin's result.

Studies of the nucleation rate in the two dimensional Ising model have
been made with the Monte Carlo method, see e.g.\ 
\cite{Alford,Rikvold,RG}. We would like to add a remark on this case.
Let
\begin{equation}
Z = \sum_{\{ S \}} \exp
\left\{ K \sum_{<ij>} S_{i} S_{j} + H \sum_{i} S_{i} \right\}
\end{equation}
be the partition function for the two-dimensional Ising model on a
square lattice with lattice spacing 1. The critical coupling is
\begin{equation}
K_{c} = \frac{1}{2} \ln ( \sqrt{2} + 1 ).
\end{equation}
The quantities appearing in Eq.~(\ref{Gamma}) can be related exactly to
$K$ and $H$ in the critical region. The interface tension is given
by \cite{Onsager}
\begin{equation}
\sigma = 2 K + \ln \tanh K
\end{equation}
and the asymmetry parameter is
\begin{equation}
\eta = 2 M H,
\end{equation}
where
\begin{equation}
M = \left( 1 - [ \sinh 2K ]^{-4} \right)^{1/8}
\end{equation}
is the zero field magnetization \cite{Yang}.

The kinetic prefactor $\kappa$, mentioned in the introduction, cannot
be calculated with static methods, because it depends on the
non-equilibrium dynamics. For dynamics that can be described by a
Fokker-Planck equation, it is, however, expected to be proportional to
the negative eigenvalue $|\lambda_{0}|$ and contributes two additional
powers of the magnetic field \cite{Langer2,RG}.

%
%
\section{Summary}

Our semiclassical calculation of the nucleation rate $\Gamma$ in the
two-dimensional Landau-Ginzburg $\phi^{4}$-model confirms Voloshin's
result (\ref{V}), which was derived in the thin wall approximation. In
particular, we confirm the prefactor value $\mathcal{A}=\eta/(2\pi)$
first obtained by Kiselev and Selivanov \cite {Kiselev}, and Voloshin
\cite{Voloshin2}.

This value differs from that obtained for the two dimensional critical
Ising model \cite{R1,R2} by the numerical factor $\pi^{2}/9\approx
1.0966$. We suppose that this small discrepancy is the result of
approximations used in \cite{R1,R2}, and the prefactor value
$\mathcal{A}=\eta/(2\pi)$ is universal.
\vspace{0.5cm}

%
\newpage
\noindent
{\large \textbf{Acknowledgments}}
\vspace{0.3cm}

One of us (S.B.~R.) would like to thank the Institute of Theoretical
Physics of the University of M\"{u}nster for hospitality.
This work is supported by the Deutsche Forschungsgemeinschaft (DFG)
under grant GRK 247/2-99 and by the Fund of Fundamental Investigations
of the Republic of Belarus.


%
\end{document}